\documentclass[pra,showpacs,twocolumn,superscriptaddress]{revtex4}%
\usepackage{color}
\usepackage{amsfonts}
\usepackage{amsmath}
\usepackage{amssymb}
\usepackage{mathrsfs}
\usepackage{graphicx}%
\providecommand{\U}[1]{\protect\rule{.1in}{.1in}}

\begin{document} 
\title{Geometric phase gate for entangling two Bose-Einstein condensates}
\author{Mahmood Irtiza Hussain}
\affiliation{State Key Laboratory of Advanced Optical Communication Systems and
Networks, School of Electronics Engineering and Computer Science, Peking
University, Beijing 100871, People's Republic of China}
\affiliation{Centre for Quantum Dynamics, Griffith University, Nathan, Queensland 4111,
Australia}

\author{Ebubechukwu O. Ilo-Okeke} 
\affiliation{Department of Physics, Federal University of Technology, P. M. B. 1526, Owerri, Imo State, Nigeria}
\affiliation{National Institute of Informatics, 2-1-2 Hitotsubashi, Chiyoda-Ku, Tokyo
101-8430, Japan}

\author{Tim Byrnes}
\affiliation{National Institute of Informatics, 2-1-2 Hitotsubashi, Chiyoda-Ku, Tokyo
101-8430, Japan}

\begin{abstract}
We propose a method of entangling two spinor Bose-Einstein condensates using a
geometric phase gate.  The scheme relies upon only the ac Stark shift and a common controllable optical mode coupled to the spins.  
Our scheme allows for the creation of an $S^{z}S^{z}$ type interaction where $S^{z} $ is the total spin. 
The geometric phase gate can be executed in times of the order of $ 2 \pi \hbar /G $, where
$ G $ is the magnitude of the Stark shift. In contrast to related schemes which relied on a fourth order interaction to produce entanglement, this is a second order interaction in the number of atomic transitions. Closed expressions for the entangling phase are derived and the effects of decoherence due to cavity decay, spontaneous emission, and incomplete de-entangling of the light to the BEC are analyzed.
\end{abstract}

\date{\today}

\pacs{03.67.Lx,67.85.Hj,03.75.Gg}
\maketitle

\section{Introduction}

The field of quantum information processing (QIP) promises the next generation
of computing technology, where fundamental logic is governed by quantum
mechanics, rather than classical physics \cite{2, 3, 4, 5, 6}. Currently, the
challenges for building a scalable QIP device are considerable, with a large
variety of systems being proposed for the task \cite{11, 12, 2, 14, 15, 16,
17, 18, 19, 20}. The difficulty lies in current limitations in the
simultaneous production of robust long-lived quantum correlations in quantum
particles while producing a scalable system \cite{7c, 8c, 9c, 10c}. At the
most elementary level, such quantum correlations require the reliable
execution of two qubit quantum gates. Such two qubit quantum gates have been
examined in several model situations involving quantum dots, microwave
potentials, nuclear spin system, superconducting qubits, trapped ions under
state dependent forces, to name several examples \cite{21, 22, 23, 24, 25, 26,
27, 28, 29, 30, 31}.

\begin{figure}[t]
\scalebox{0.3}{\includegraphics{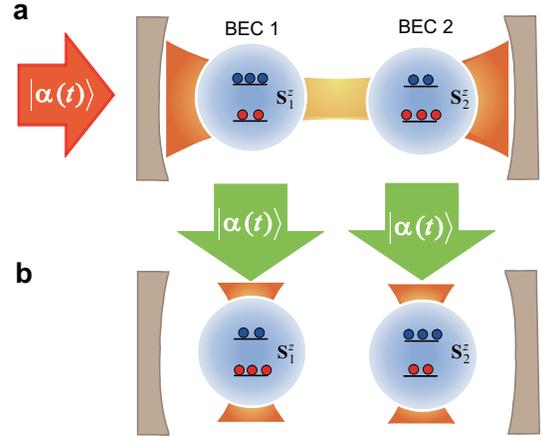}}
\caption{Schematic experimental configuration for the geometric phase gate.  Two macroscopic spins (shown as BECs here) are
placed in a cavity such that an ac Stark shift occurs on the energy levels. The geometric phase gate then performed by 
the following procedure.  (a) A laser is applied to the cavity such that both BECs are illuminated, and controlled such as to 
follow the evolution $ | \alpha(t) \rangle $ for a time $ t = [0,T] $.  The phase (\ref{halfwayphase}) is induced at this point. 
(b) The laser is then applied individually to each BEC, with the opposite detuning $ \Delta' = - \Delta $ following the same 
displacement force $ F(t) $.  This completes the geometric phase gate and a $ S^z S^z $ interaction is induced between the BECs.}
\label{fig1}
\end{figure}

For macroscopic objects such as Bose-Einstein condensates (BECs), proposed
schemes for entanglement generation are less developed. The only proposed
scheme for BECs, to our knowledge, involves a photon mediated scheme for BECs placed in
optical cavities \cite{byrnes12,pyrkov13}.  Other possible methods include those originally
formulated for single atoms, using state-dependent forces \cite{31}. Experimentally, entanglement
between a BEC and a single atom was achieved in Ref. \cite{lettner11}. For
atomic ensembles, entanglement and teleportation has been performed using a
continuous variable approach \cite{julsgaard01,krauter12}. Here, the
entanglement is in the form of two-mode squeezing in the total spin variables
of the ensembles. Another approach involves using spin wave states, where
teleportation was recently achieved \cite{bao12}. Besides these, geometric
phase gate approach first introduced for trapped ions \cite{29,28}, has been
shown to be a robust and fast method of creating entanglement between two
qubits \cite{leibfried03}. 
The geometric phase gate is an attractive method 
for producing an entangling gate from the point of view
of its robustness, as it may tolerate various imperfections such as variations 
in the initial conditions of the common bosonic mode \cite{28}. Here we apply the geometric phase to control 
the state of light in phase-space to achieve a fast and robust $S^zS^z$ interaction 
between two different collective spins. We show in this paper that by using the geometric phase gate, entanglement is possible using the ac Stark shift Hamiltonian, which is a second order process. This improves upon previous works using photon-mediated entanglement 
such as Ref. \cite{pyrkov13} which relied on a fourth order transition to produce the entangling gate.

Creating entanglement between collective spin states, apart from the perspective of 
fundamental interest in macroscopic entanglement \cite{vedral08}, is a key element to
QIP based on spin coherent states. Previously we have introduced a scheme of
performing quantum information processing using macroscopic states encoded on
two-component BECs \cite{byrnes12}. The basic idea involves using spin
coherent states in place of genuine qubits. It has been shown that many QIP
schemes such as quantum algorithms \cite{byrnes12}, quantum teleportation
\cite{pyrkov13b}, and quantum communication \cite{34} can be performed using
such ``BEC qubits''. In order to perform such QIP protocols, it is necessary to
produce entanglement between different BEC qubits. Similarly to standard
qubits, for universality the scheme requires at least one and two BEC qubit
control \cite{byrnes12}. As coherent control of single two-component BEC has
already been achieved \cite{bohi09}, a major remaining technological issue is in
the efficient creation of two BEC entanglement. We note that the structure of
the entanglement between two spinor BECs has been recently predicted to
display a fractal entanglement (the ``devil's crevasse'') \cite{byrnes13}, thus displays
interesting physics in its own right.

\section{Geometric Phase Gate}

\subsection{The Protocol}

In order to realize the geometric phase gate for the collective spins mediated by cavity photons, we consider the following Hamiltonian 
(see Appendix for a derivation)
\begin{align}
H    = & \hbar\omega_{0}a^{\dag}a+G(S_{1}^{z}+S_{2}^{z})a^{\dag} a \nonumber\\
&  - \frac{F(t)}{\sqrt{2}} \left(  ae^{i(\omega_{0}-\Delta)t}+a^{\dag
}e^{-i(\omega_{0}-\Delta)t} \right)  , \label{BECH}%
\end{align}
where $a$ is an annihilation operator for a common bosonic mode, $S^{z} $ is the total $z$-spin of the BEC. The first term in (\ref{BECH}) gives the energy of the bosonic mode $\hbar\omega_0$, the
second is an ac Stark shift to the spin states, and the last term is the displacement operator for the bosonic mode that is controllable with a time-dependent coefficient $F(t)$.

The basic idea of the scheme is presented in Figure \ref{fig1}. The two BECs are placed in a cavity and illuminated with the same controllable laser field of the form of a coherent state $| \alpha(t) \rangle$ \cite{zheng04}. The laser
is far detuned from the transition to an excited state of the qubit, such that
there is an ac Stark shift to the energy level, giving rise to the second term
in (\ref{BECH}). Initially the cavity starts in the vacuum state $|\alpha(t=0) \rangle= |0 \rangle$, and it is controlled by the displacement operation such that after a time $ t=T $ it returns to its original state.  After this point the BECs become entangled due to a spin-dependent geometric phase. Let us illustrate the general procedure by taking the example of qubits instead of BEC.  Assuming the particular case of initially $ x $-polarized spins, the combined state of the system is
\begin{align}
\frac{1}{2} |0 \rangle ( | \uparrow_1  \rangle +| \downarrow_1  \rangle ) (| \uparrow_2  \rangle +  | \downarrow_2  \rangle )
\end{align}
where $ | \sigma_i \rangle $ with $ \sigma = \uparrow, \downarrow $ and $ i = 1,2 $ are the basis states of the two qubits.  The light is then controlled through phase space $(\mbox{Re} (\alpha), \mbox{Im} (\alpha)) $ in a state dependent way such that partway through the evolution the light and the qubits become entangled with light
\begin{align}
& \frac{1}{2} \Big( e^{i \Phi_{\uparrow \uparrow } (t)} |\alpha_{\uparrow \uparrow } (t)  \rangle  | \uparrow_1 \uparrow_2  \rangle
+ e^{i \Phi_{\uparrow \downarrow }(t)} |\alpha_{\uparrow \downarrow } (t)\rangle  | \uparrow_1 \downarrow_2  \rangle \nonumber \\
& + e^{i \Phi_{\downarrow \uparrow }(t)} |\alpha_{\downarrow \uparrow }(t) \rangle  | \downarrow_1 \uparrow_2  \rangle
+ e^{i \Phi_{\downarrow \downarrow }(t)} |\alpha_{\downarrow \downarrow } (t)\rangle  | \downarrow_1  \downarrow_2  \rangle \Big),
\label{imperfect}
\end{align}
where $ \Phi $ is a spin-dependent phase picked up due to the evolution of the light. 
By demanding that the state of light at the end of the evolution is the same as the initial state for all terms in the superposition, we then have
\begin{align}
& \frac{1}{2} |0 \rangle  \Big( e^{i \Phi_{\uparrow \uparrow } (T)  }  | \uparrow_1 \uparrow_2  \rangle
+ e^{i \Phi_{\uparrow \downarrow } (T) }  | \uparrow_1 \downarrow_2  \rangle \nonumber \\
& + e^{i \Phi_{\downarrow \uparrow } (T) }  | \downarrow_1 \uparrow_2  \rangle
+ e^{i \Phi_{\downarrow \downarrow } (T)}  | \downarrow_1  \downarrow_2  \rangle \Big),
\end{align}
which is for suitable $ \Phi $ an entangled state. The procedure for the BEC is similar, but now the spins are expanded in 
terms of $ S^z $, taking eigenvalues $ [-N,-N+2,\dots,N] $ \cite{byrnes12}. 

In standard derivations of the geometric phase such as that given in Ref. \cite{28}, 
state-dependent forces are used to create the state dependence to the phase $\Phi_{S^z_1 S^z_2} (T) $, thus 
the spin-boson interaction comes in the last term of (\ref{BECH}), rather than the ac Stark shift
as we have here.  Nevertheless, we will see that this creates a state-dependent phase 
$\Phi_{S^z_1 S^z_2} (T) $, thus creating entanglement. The method based on the ac Stark shift may also be used for standard qubits \cite{hussain13,30}, although we shall
concentrate upon the macroscopic spin case in this paper.

\subsection{Entangling phase}

We assume that the state of the BECs are initially unentangled, such that the initial state is 
$ \sum_{S^z_1 S^z_2 } \psi_1(S^z_1) \psi_2 (S^z_2) | S^z_1 S^z_2 \rangle $. The total state of the light and the atoms 
then follows the ansatz
\begin{align}
|\psi(t) \rangle = \sum_{S^z_1 S^z_2 }  e^{i \Phi_{S^z_1 S^z_2}  (t)} | \alpha_{S^z_1 S^z_2} (t) \rangle   \psi_1(S^z_1) \psi_2 (S^z_2)  | S^z_1 S^z_2 \rangle .
\end{align}
Substituting into the evolution equation $ i \hbar d | \psi(t) \rangle/dt = H | \psi(t) \rangle $ we obtain (see Appendix for details)
\begin{equation}
\label{eq:pap06}
\begin{split}
\dot{\alpha}_c& =  i\frac{F(t)}{\hbar\sqrt{2}}e^{i\Omega t},\\
\dot{\Phi} & = -\frac{i}{2}\left(\dot{\alpha}_c\alpha^*_c - \alpha_c\dot{\alpha}^*_c \right) 
\end{split}
\end{equation} 
where
\begin{equation}
\label{omegarelation}
\Omega =\Delta+\frac{G(S_{1}^{z}+S_{2}^{z})}{\hbar},
\end{equation}
$\alpha_c = \alpha e^{i[\omega_0 + \tfrac{G}{\hbar}(S_1^z + S_2^z)]t}$, and we have dropped the $ S^z_1 S^z_2 $ labels on $ \alpha $ and $ \Phi $ for brevity.  Starting from an initial amplitude $\alpha_c(0) = 0$, the above
creates a time dependent displacement according to
\begin{equation}
\alpha_c(t)=  \frac{i}{\hbar\sqrt{2}}\int\limits_{0}^{t} d\tau\, F(\tau)e^{i\Omega \tau}
\label{alphacevol}
\end{equation}
We impose that after a time $T$, the coherent state returns back to its
original state. This requires the condition
\begin{align}
\int_{0}^{T}d\tau \, F(\tau) e^{i \Omega \tau}=0, \label{conditionbec}%
\end{align}
The phase~\cite{29,28} picked up by the coherent state is then 
\begin{equation}
\Phi \left(  T\right) =  \frac{1}{2\hbar^2}\mathrm{Im} \int\limits_{0}^{T}d\tau_1\, \int\limits_{0}^{\tau_1} d\tau_2\,F(\tau_1) F(\tau_2) e^{i\Omega \tau_r}  , \label{becphase}%
\end{equation}
where $\tau_{r}=\tau_{1}-\tau_{2}$.  To show that this is in fact an entangling gate, we
now expand the exponential in $ G/\hbar $ using a Taylor series up to second order. We later
give a concrete example of how (\ref{conditionbec}) may be satisfied, while justifying the expansion. The phase we obtain is 
\begin{equation}
\begin{split}
\Phi(T) & = \phi_{0} (\Delta) +\phi_{1}(\Delta) (S_{1}^{z}+S_{2}^{z})
\\
& +\frac{\phi_{2}(\Delta)}{2} \left[ (S_{1}%
^{z})^{2}+(S_{2}^{z})^{2} \right] 
+\phi_{2}(\Delta) S_{1}^{z}S_{2}^{z} \label{halfwayphase}%
\end{split}
\end{equation}
where
\begin{align}
\phi_{0}(\Delta)  &  =\frac{1}{2\hbar^{2}}\int_{0}^{T}d\tau_{1} \int_{0}^{\tau_{1}%
}d\tau_{2} \sin(\Delta\tau_{r}) F\left(  \tau_{1}\right)  F\left(  \tau
_{2}\right)  ,\nonumber\\
\phi_{1}(\Delta)  &  =\frac{G}{2\hbar^{3}}\int_{0}^{T}d\tau_{1} \int
_{0}^{\tau_{1}}d\tau_{2}\tau_{r} \cos(\Delta\tau_{r})  F\left(  \tau_{1}\right) F\left(  \tau_{2}\right)  ,\nonumber\\
\phi_{2}(\Delta)  &  =-\frac{G^2}{2\hbar^{4}}%
\int_{0}^{T}d\tau_{1} \int_{0}^{\tau_{1}}d\tau_{2}\tau_{r}^{2} \sin(\Delta
\tau_{r}) F\left(  \tau_{1}\right)  F\left(  \tau_{2}\right)  .
\end{align}
We see that the phase is proportional to
$S_{1}^{z}S_{2}^{z}$, which is the entangling operation as desired. There are
however also terms of the form $(S_{i}^{z})^{2}$, which would not be present for qubits 
since $(S^{z}/N)^{2}\neq I$ in contrast to Pauli operators.
Hamiltonians of this form correspond to squeezing operators \cite{riedel10}
and can be of use in quantum metrology applications \cite{sorensen01}.
However, for our purposes this is an unwanted by-product of the geometric
phase gate and require elimination using a suitable procedure.

\begin{figure}
\scalebox{0.3}{\includegraphics{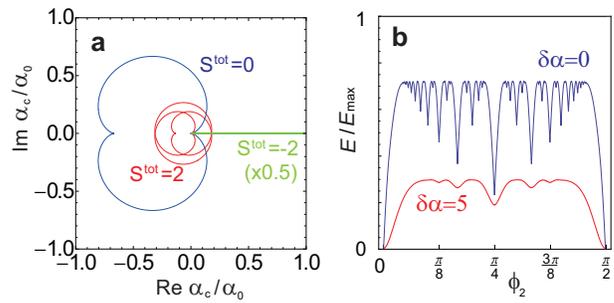}}
\caption{\label{fig2}
(a) Various trajectories of the coherent states during the evolution for various total spins $ S^{\mbox{\tiny tot}} = S^z_1 + S^z_2 $. Parameters used are $ m = 1 $ and $ n = 1 $ and the coherent states are normalized to $ \alpha_0 = \frac{i F_0}{\sqrt{2} G} $. (b) The entanglement as measured by the logarithmic negativity $ E = \log_2 || \rho^{T_2} || $ normalized to the maximal entanglement $ E_{\mbox{\tiny max}} = \log_2 (N+1) $. The initial states of the two BECs are assumed to be in maximum $S^x_i $ eigenstates, $ i = 1,2$. The ideal case of $ \delta \alpha =0 $ is shown for comparison. The number of bosons $ N $ in the BEC is taken to be $ N = 20 $ and $ \delta \theta = 0.1 $.}
\end{figure}

\subsection{Elimination of undesired phases}


To remove the undesired terms in (\ref{halfwayphase}) we subject our system
with another set of laser pulses obeying the Hamiltonian
\begin{align}
H^{\prime}=  &  \sum_{i=1,2} \hbar\omega_{0}a_{i}^{\dag}a_{i}+GS_{i}%
^{z}a_{i}^{\dag}a_{i}  \nonumber\\
&  -\frac{F(t)}{\sqrt{2}} \left(  a_{i} e^{i(\omega_{0}-\Delta^{\prime}%
)t}+a_{i}^{\dag}e^{-i(\omega_{0}-\Delta^{\prime})t} \right)  , \label{BECH2}%
\end{align}
where $a_{i} $ refers to the photons in cavity $i =1,2 $. Such a Hamiltonian
may be performed by separately illuminating the BECs (Figure \ref{fig1}b). 
Performing the same calculation, the phase that is picked up by each BEC $ i =1,2 $ is then
\begin{align}
\Phi^{\prime}_{i}(T)=\phi_{0}(\Delta') +\phi_{1}(\Delta')  S_{i}^{z}+\frac
{\phi_{2}(\Delta') }{2}(S_{i}^{z})^{2} \label{halfwayphase2}.
\end{align}
The phase for each BEC is precisely the same as that in
(\ref{halfwayphase}) except that the $S^{z}_{1} S^{z}_{2} $ is missing.
Therefore by applying the reverse phase $\phi_{2} (\Delta')= - \phi_{2} (\Delta)$ we
may eliminate the undesired terms in (\ref{halfwayphase}). This may be achieved by choosing $\Delta^{\prime}= - \Delta$. 
The total phase after the two operations is
\begin{align}
\Phi_{\mbox{\tiny tot}}  &  =\Phi+\Phi^{\prime}_{1} + \Phi^{\prime}%
_{2}\nonumber\\
&  = 2 \phi_{1} ( S^{z}_{1} + S^{z}_{2} ) + \phi_{2} S^{z}_{1} S^{z}_{2} .
\end{align}
Our final expression for the phase contains the desired $S_{1}^{z}S_{2}^{z}$
interaction, together with a single qubit rotation term which may be
compensated for using single BEC qubit control. We note that a similar method was 
presented in Ref. \cite{sorensen00} to create $ (S^z)^2 $ interactions.

\section{Example solution}



We now give an example solution for satisfying (\ref{conditionbec}).
Unlike the original formulation for geometric phase gates in Ref. \cite{29,28}, in our case $\Omega$ is state dependent, so that (\ref{conditionbec}) must be satisfied for all possible $ | S^z_1 S^z_2 \rangle $ states.  Choosing
\begin{align}
\hbar\Delta=2Gn
\label{detuning}
\end{align}
where $n$ is an integer, the set of frequencies that $\Omega$ take are then
\begin{align}
\hbar\Omega=2G(n-N),2G(n-N+2),\dots,2G(n+N),
\end{align}
where $ N $ is the maximal eigenvalue of $ S^z $.  Noting that all the frequencies are even multiples of $ G/\hbar $, Eq. (\ref{conditionbec}) may then be satisfied by choosing
\begin{align}
F(\tau)  &  =F_0\sin(Gm\tau/\hbar) \nonumber\\
T  &  =\frac{2\pi\hbar}{G}%
\label{fandt}
\end{align}
where $m$ is an odd integer. The phases induced by the geometric phase gate
may then be evaluated exactly to give
\begin{align}
\phi_0 & = -\frac{\pi F_0^2  n}{G^2 (m^2 - 4 n^2)} \nonumber \\
\phi_{1}  & =- \frac{\pi F_{0}^{2} \left(  3m^{2}+4n^{2}\right) }{2G^{2}\left(  m^{2}-4n^{2}\right)
^{2}} ,\nonumber  \\
\phi_{2}  & =-\frac{ 2\pi F_{0}^{2} n\left(  11m^{2}+4n^{2}\right)   }{G^{2}\left(  m^{2}-4n^{2}\right)  ^{3}%
}. \label{phi2plot}
\end{align}
We observe that for $ n \gg m $, the phases decrease in steps of $n $. Accounting for the fact that $ S^z \sim O(N) $, the expansion Eq. (\ref{halfwayphase}) may be made to converge if $ n \sim N $.  The remaining parameters $ F_0 $ and $ m $ may 
then be used to tune the phase to the desired value. The decrease of the coefficients $ \phi_i $ justifies the expansion 
made in (\ref{becphase}) thus showing that to a good approximation $ S^z S^z $ interactions can be implemented while
exactly satisfying (\ref{conditionbec}).

In an realistic experimental situation, the control of the coherent fields $ | \alpha (t) \rangle $ will not be perfect, which can arise due to a variety of reasons such as technical noise and cavity decay. The imperfect control will leave the coherent state with a remnant component $ \alpha (T) \ne 0 $.   In general this will lead to imperfect disentangling, leading to decoherence. Figure \ref{fig2}(a) show various trajectories of the coherent states for various spin states as evolved by (\ref{alphacevol}) and 
using (\ref{fandt}).  Various $ S^z_1 + S^z_2 $ eigenstates follow different trajectories, leaving the final state after time $ T $ with a random remnant offset $ \alpha_c (T) \ne 0 $.  This effect may be simply modeled by assuming that the $ | S^z_1 S^z_2 \rangle $ basis state of the BEC becomes entangled with the coherent state $ | \delta \alpha e^{-i(S^z_1 + S^z_2)\delta \theta} \rangle $, where $ \delta \alpha $ is the remnant offset at the end of the evolution, and the phase variation $ \delta \theta $ arises due to the conversion from $ \alpha_c $ to $ \alpha $. The dependence of the imperfect disentangling on the entanglement may be calculated by finding the reduced density matrix $ \rho = \mbox{Tr}_a |\psi \rangle \langle \psi |  $ where  $ | \psi \rangle $ is the entangled state of the BECs and the photons after the evolution, and the trace is over the photon degrees of freedom.
Figure \ref{fig2}(b) shows the logarithmic negativity  \cite{vidal02} of the state following the geometric phase gate and shows a diminished entanglement as expected.  The effect of the imperfect disentangling is similar to a $ S^z $-dephasing by diminishing the off-diagonal components of the density matrix.

\section{Experimental parameter estimation}

We now give some experimental details of our scheme. 
The displacement terms in
(\ref{BECH}) and (\ref{BECH2}) which realize the geometric path of the laser
in phase space can be performed using standard quantum optical methods \cite{bachor04}. We thus
describe the atomic configuration that would realize the Hamiltonians, specializing to the BEC case. The quantum information is stored in the hyperfine ground states of the BEC. For a two-component BEC implementation as given in \cite{bohi09}, states creating the spinor are 
 $ | F=1, F^z=-1 \rangle $ and $ | F=2, F^z=1 \rangle $.  A circularly polarized laser pulse is detuned from the atomic resonance transition is incident on the atoms~\cite{49,50}. The beam couples one of the ground
states to an excited state giving rise to an ac Stark shift as given in the second term of (\ref{BECH}).  

There are two primary sources of external decoherence in our scheme: spontaneous emission and cavity photon loss. Considering spontaneous emission first, the ac Stark shift involves a virtual excitation to an excited
state, which is susceptible to spontaneous decay.  For a effective coupling $ G = \frac{g_0^2}{\Delta }  $ the effective 
decoherence rate is $ \Gamma_{\mbox{\tiny eff}} =  \frac{\Gamma N  g_0^2 }{\Delta^2} $ \cite{pyrkov13}. Here $ g_0 $ is the single atom cavity coupling, $\Delta $ is the detuning, and $ \Gamma $ is the spontaneous emission rate. The factor of $ N $ arises due to stimulated emission of the excited state into the ground states.  In order that spontaneous emission does not cause the state of the BEC to decohere, we require that the time required for the operation is within the effective decoherence time  $ 1/ \Gamma_{\mbox{\tiny eff}} $. We thus require 
\begin{align}
\frac{\hbar}{G} \le \frac{\Delta^2}{\Gamma N g_0^2} 
\end{align}
which gives the first constraint on the detuning that should be used
\begin{align}
\Delta \ge \hbar \Gamma N  . 
\label{firstconstraint}
\end{align} 

Cavity photon loss will affect our scheme as during the evolution there are superpositions of coherent states which depend upon the spin state, as illustrated in Fig. \ref{fig2}(a).  While a coherent state $ | \alpha \rangle $ remains a pure state even in the presence of loss, superpositions of coherent states such as $ | \alpha \rangle \pm  | - \alpha \rangle $ decohere into a mixture with an effective
rate $ \kappa_{\mbox{\tiny eff}} =  \kappa  |\alpha|^2 $ \cite{walls85,savage85,goetsch95}. In our case the magnitude of the coherent states is determined by $ \alpha \sim F_0/G $ as $ F_0/\hbar $ is the rate of displacement  and $ \hbar/G $ is the time of the evolution.  We thus have a second constraint such that the coherent state superposition does not decohere during the evolution
\begin{align}
\frac{\hbar}{G} \le \frac{1}{\kappa |\alpha|^2}.  
\end{align}
Combining this with (\ref{firstconstraint}) we obtain a constraint on the brightness of the coherent light that can be used
\begin{align}
|\alpha |^2 \sim \frac{F_0^2}{G^2} \le \frac{g_0^2}{\hbar^2 \kappa \Gamma N } .
\label{secondconstr}
\end{align}

Let us now see how these contraints compare with experimental parameters. The cavity-BEC coupling may be estimated using parameters in Ref. \cite{colombe07} and is $ g_0/\hbar = 1350 \mbox{MHz} $ with a cavity decay rate of $ \kappa = 330 \mbox{MHz} $ and a spontaneous 
emission rate of $ \Gamma = 19 \mbox{MHz} $.  Then for example, the ac Stark shift 
assuming a detuning (\ref{firstconstraint}) and $ N = 10^3 $ is $ G/\hbar \sim  \Gamma_{\mbox{\tiny eff}} \approx 96 \mbox{MHz} $, for a detuning of $ \Delta/\hbar  = 19  \mbox{GHz} $.  The second constraint (\ref{secondconstr}) then evaluates to $ |\alpha|^2 \le 0.3 $, which corresponds to rather weak coherent light pulses, but within experimental feasibility.  For smaller atom numbers the estimates improve as  (\ref{firstconstraint}) means that smaller detunings can be used, increasing $ G $.  This in turn increases the photon number in (\ref{secondconstr}) which allows for larger phases to be generated in (\ref{phi2plot}).  Naturally, as the quality of the cavities improve (\ref{secondconstr}) allows also for brighter coherent states to be used.

\section{Conclusions}

In summary, we investigated a scheme for entangling two macroscopic spin states via a geometric
phase gate. Rather than a state dependent force used in the original ion trap formulation of the 
gate, the scheme is based on the ac Stark shift which is routinely realized in experiments.  
One difference between the the original formulation and the present work
is in the condition (\ref{conditionbec}) that must be satisfied in order for the coherent
state to be disentangled at the end of the displacement operation.  Due to the spin dependent 
frequency $ \Omega $, this gives a more complicated condition that needs to 
be satisfied. Despite this, by choosing a suitable form of the force $ F(t) $, this may be 
satisfied exactly.  The timescale of the geometric phase gate is determined by the ac Stark shift coupling $ G $  which 
in turn is dependent on the cavity-atom coupling and the detuning. 
The main decoherence mechanisms are cavity photon decay, which provides the common mode for the BECs to entangle with, and spontaneous emission of the atoms. 
It is likely possible to improve upon the gate time to decoherence time ratio, using alternative
solutions to that presented here satisfying (\ref{conditionbec}).  For example, 
using very high frequency $ F(t) $ would approximately satisfy (\ref{conditionbec}) as long as frequencies above $ \sim GN/\hbar $ are used. This would allow for much shorter gate times, at the price of a larger $ F_0 $. The many possible solutions for the geometric phase gate is one of the reasons for the 
robustness of the approach, which gives it a degree of tolerance for experimental imperfections as the phase $ \Phi $ is determined
by the trajectory on the phase space of the light. Due to spontaneous emission scaling with the number of atoms in the BEC $ N $, currently the scheme is limited to relatively small BECs, around $ N = 10^3 $. For BECs with small numbers of atoms, the proposed method 
is viable with current experimental parameters. As the qualities of cavities improve, this would allow for the reliable production of entanglement between macroscopic objects, which could be employed for various quantum information tasks.

\section*{ACKNOWLEDGMENTS}

This work is supported by the Transdisciplinary Research Integration Center,
the Okawa Foundation, National Science Fund for Distinguished Young Scholars
of China (Grant No. 61225003), National Natural Science Foundation of China
(Grant No. 61101081), and the National Hi-Tech Research and Development (863) Program, the Inamori Foundation, 
NTT Basic Laboratories, and JSPS KAKENHI Grant Number 26790061.

\appendix

\section{ac Stark shift Hamiltonian}
Consider a Bose-Einstein condensate (BEC) placed in a cavity that is detuned from atomic resonance transition, and coupling the ground state to an excited state of the atom. 
The Hamiltonian describing the composite system is given by
\begin{equation}
\label{eq:app01}
\hat{H} = \hat{H}_{\mbox{\tiny atom}}+ \hat{H}_{\mbox{\tiny light}} + \hat{H}_{\mbox{\tiny atom-light}},
\end{equation}
where 
\begin{equation}
\label{eq:app02}
\begin{split}
& \hat{H}_{\mbox{\tiny atom}} = \sum_j \int d\mathbf{r} \Big[\hat{\psi}_{gj}^\dagger(\mathbf{r}) H_0 \hat{\psi}_{gj}(\mathbf{r}) + \hat{\psi}_{ej}^\dagger(\mathbf{r}) H_0 \hat{\psi}_{ej} (\mathbf{r})\Big], \\  
& \hat{H}_{\mbox{\tiny light}} =\frac{ \hbar\omega}{2}\left(\hat{a}^\dagger \hat{a} + \hat{a}\hat{a}^\dagger\right), \\
& \hat{H}_{\mbox{\tiny atom-light}}  = - \sum_j \int d\mathbf{r} \,\hat{\psi}_{gj}^\dagger(\mathbf{r})\mathbf{d}_j \cdot\hat{\mathbf{E}}(\mathbf{r},t) \hat{\psi}_{ej}(\mathbf{r})+ \mbox{H.c.}
\end{split}
\end{equation}
and 
\begin{equation}
H_0 = - \frac{ \hbar^2}{2m} \nabla^2 + V (\mathbf{r}) 
\end{equation}
is the single-particle Hamiltonian including the confining potential $  V (\mathbf{r}) $ of the atoms with mass $ m$. The indices $g,e$ label the ground and excited states of the atom respectively. We consider that the atoms may possess several types of ground and excited states, such as hyperfine states, which are labeled by the index $ j $.  We will consider the case of two-component BECs, hence for our case the sum runs over $ j =1,2 $. 
The bosonic field operators $\hat{\psi}^\dagger (\mathbf{r}) $ acting on vacuum state create an atom at the position $ \mathbf{r} $.  For the photons we work in the single mode approximation where we consider only the 
cavity photon mode, with $a$ being the photon annihilation operator.  $\mathbf{d}_j $ is the dipole moment of the atom. In the single mode approximation the electric field operator is 
\begin{equation}
\hat{\mathbf{E}}(\mathbf{r},t) = \hat{a}{{E_-(\mathbf{r})}}\mathbf{\varepsilon}  e^{-i\omega t} + \hat{a}^\dagger{E_+(\mathbf{r})}\mathbf{\varepsilon}^*  e^{i\omega t},
\end{equation}
 where $\mathbf{\varepsilon} $ is a complex unit vector describing the polarisation of light, $\omega$ is the frequency of the incident light, $E_\pm(\mathbf{r}) $ is the complex amplitude of electric field.

The atom-light interaction Hamiltonian consist of terms that are energy conserving and non-conserving energy terms. Making rotating-wave approximation in order to remove the energy non-conserving terms gives
\begin{equation}
\label{eq:app03}
\begin{split}
\hat{H}_{\mbox{\tiny atom-light}}^{RW} & = - \sum_j \int d\mathbf{r} \hat{\psi}_g^\dagger(\mathbf{r})\mathbf{d}_j \cdot\hat{\mathbf{E}}^{(+)} e^{i\omega t} \hat{\psi}_e(\mathbf{r}) + \mbox{H.c.} 
\end{split}
\end{equation}
where $\hat{\mathbf{E}}^{(+)} = \hat{a}^\dagger {{E_+(\mathbf{r})}}\mathbf{\varepsilon}^* $. In the regime of very strong detuning, the excited states of the atoms are hardly populated and therefore can be eliminated adiabatically. The effective atom-light interaction Hamiltonian as seen by the atoms in the ground state is 
\begin{equation}
\label{eq:app05}
\hat{H}_{\mbox{\tiny eff}}^{\mbox{\tiny int}} = -2 \sum_j \int d\mathbf{r}\, \hat{\psi}^\dagger_{gj} (\mathbf{r})\,\frac{\mathbf{d}_j \cdot\hat{\mathbf{E}}^{(+)}\mathbf{d}_j \cdot\hat{\mathbf{E}}^{(-)}}{\hbar\Delta_j}\,\hat{\psi}_{gj} (\mathbf{r}),
\end{equation}
where $\Delta_j= \omega_{ej} - \omega_{gj} - \omega$, and $\mathbf{d}_j \cdot\hat{\mathbf{E}}^{(-)} =E_-(\mathbf{r}) \langle e j |\mathbf{d}\cdot\varepsilon|g j\rangle \hat{a}$ are the dipole transition matrix elements. Assuming all the atoms are in the ground state of the trapping potential, we may write  
\begin{equation}
\hat{\psi}^\dagger_{gj}(\mathbf{r}) = b^\dagger_{gj} \psi_{g}^*(\mathbf{r})
\end{equation}
where $\psi_g(\mathbf{r})$ is the wave function of an atom in the ground state (we assume this is independent of $ j$) and $b^\dagger_{gj}$ is an operator that acts on vacuum state to create an atom in the ground state. For a two component BEC, the effective interaction Hamiltonian can be written in terms of their relative population difference as 
\begin{equation}
\label{eq:app06}
\hat{H}_{\mbox{\tiny eff}}^{\mbox{\tiny int}} = -\left[G(b^\dagger_1 b_1 - b^\dagger_2 b_2) + g\hat{N}\right]\hat{a}^\dagger\hat{a},
\end{equation}
where $\hat{N} =b^\dagger_1 b_1 +  b^\dagger_2 b_2$ is the total atom number operator, $G = G_1 - G_2$ and $g = G_1 + G_2$ with 
\begin{equation}
\label{eq:app07}
G_j = \int\,d\mathbf{r} |E_-(\mathbf{r})|^2\,  \frac{ \psi^*(\mathbf{r})\langle g j |\mathbf{d}_j \cdot\varepsilon^*|e j \rangle \langle e j |\mathbf{d}_j \cdot\varepsilon|g j\rangle \psi(\mathbf{r})}{\hbar\Delta_j}, 
\end{equation}
being  the strength of the atom-light interaction experienced by the \emph{j}th ground state. Rewriting (\ref{eq:app06}) in terms of  $S^z = b^\dagger_1b_1 -  b^\dagger_2 b_2$, gives the ac Stark  Hamiltonian.

\section{Derivation of the evolution equations for the coherent state}

The time evolution of the state $ | \psi(t) \rangle $ is determined by
\begin{align}
& i \hbar \left[ i \dot{\Phi} | \gamma \rangle  - \left( \frac{ \dot{\alpha} \alpha^* + \alpha \dot{\alpha^*}}{2} \right) | \gamma \rangle + \dot{\alpha} a^\dagger | \gamma \rangle \right] = \nonumber \\
& \left[ \hbar \omega_0 \alpha + \hbar G (S^z_1 + S^z_2) \alpha - \frac{F(t) e^{-i(\omega_0 - \Delta)t} }{\sqrt{2}} \right] a^\dagger | \gamma \rangle \nonumber \\
& - \frac{F(t) \alpha e^{i(\omega_0 - \Delta)t}}{\sqrt{2}} | \gamma \rangle
\label{timeeveq}
\end{align}
where $ | \gamma \rangle = e^{i \Phi(t)} | \alpha(t) \rangle $.  Now decompose the state $ a^\dagger | \gamma \rangle $ into $ | \gamma \rangle $ and the state orthogonal to this $ | \gamma_\perp \rangle  $ by a Gram-Schmidt process.  We may now write $ a^\dagger | \gamma \rangle = A | \gamma \rangle + B | \gamma_\perp \rangle $.  Substituting this into (\ref{timeeveq}), we obtain the equations of motion
\begin{align}
i\hbar \left[ i \dot{\Phi} - \left( \frac{ \dot{\alpha} \alpha^* + \alpha \dot{\alpha^*}}{2} \right)  \right]= - \frac{F(t) \alpha e^{i(\omega_0 - \Delta)t}}{\sqrt{2}}  \\
\dot{\alpha} = - i \left[ \omega_0 + G (S^z_1 + S^z_2) \right] \alpha + \frac{i F(t) e^{-i(\omega_0 - \Delta)t} }{\hbar \sqrt{2} } .
\end{align}
Making the transformation into the rotating frame $\alpha_c = \alpha e^{i[\omega_0 + \tfrac{G}{\hbar}(S_1^z + S_2^z)]t}$ yields the evolution equations in the main text.


\end{document}